\documentclass[prl,twocolumn]{revtex4} 
\usepackage{graphicx}
\usepackage{amssymb,amsmath}

\begin{document}

\title{
%A  Berry-Keating  model with closed periodic orbits and smooth  Riemann spectrum
The $H=xp$ model  revisited and the   Riemann zeros
%Closing the classical orbits  in the $H=xp$ model
%The   $H=xp$ model  revisited 
}

\author{Germ\'an Sierra$^*$  and Javier Rodr\'{\i}guez-Laguna$^\dagger$}

\affiliation{$*$Instituto de F\'{\i}sica Te\'orica, CSIC-UAM, Madrid, Spain \\
$\dagger$Universidad Carlos III, Madrid, Spain}

% started the  January 2011

\bigskip\bigskip\bigskip\bigskip

%%%%%%%%%%%%%%% MATH CHARACTERS %%%%%%%%%%%%%%%%%%%%%%%%%%%%
%
\font\numbers=cmss12
%\font\numbers=cmu10 scaled\magstep1
\font\upright=cmu10 scaled\magstep1
\def\stroke{\vrule height8pt width0.4pt depth-0.1pt}
\def\topfleck{\vrule height8pt width0.5pt depth-5.9pt}
\def\botfleck{\vrule height2pt width0.5pt depth0.1pt}
\def\Zmath{\vcenter{\hbox{\numbers\rlap{\rlap{Z}\kern
0.8pt\topfleck}\kern 2.2pt
                   \rlap Z\kern 6pt\botfleck\kern 1pt}}}
\def\Qmath{\vcenter{\hbox{\upright\rlap{\rlap{Q}\kern
                   3.8pt\stroke}\phantom{Q}}}}
\def\Nmath{\vcenter{\hbox{\upright\rlap{I}\kern 1.7pt N}}}
\def\Cmath{\vcenter{\hbox{\upright\rlap{\rlap{C}\kern
                   3.8pt\stroke}\phantom{C}}}}
\def\Rmath{\vcenter{\hbox{\upright\rlap{I}\kern 1.7pt R}}}
\def\Z{\ifmmode\Zmath\else$\Zmath$\fi}
\def\Q{\ifmmode\Qmath\else$\Qmath$\fi}
\def\N{\ifmmode\Nmath\else$\Nmath$\fi}
\def\C{\ifmmode\Cmath\else$\Cmath$\fi}
\def\R{\ifmmode\Rmath\else$\Rmath$\fi}
\def\H{{\cal H}}
\def\NN{{\cal N}}
\def\cl{{\rm cl}}
\def\RS{{\rm RS}}
\def\E{{\rm E}}
\def\b{{\bf b}}
\def\tpsi{\tilde{\psi}} 
\def\bS{{\bf S}}

\begin{abstract}
Berry and Keating conjectured   that the classical Hamiltonian $H = xp$ is related to the  Riemann zeros.
A regularization of  this model yields  semiclassical energies  that behave, in average,  as the non trivial zeros of the Riemann zeta function.
However,  the classical trajectories are not closed, rendering   the  model incomplete.
%leads to inconsistencies in the classical and quantum theories.
 In this paper, 
we show that the Hamiltonian $H = x (p +  \ell_p^2/p)$  contains closed periodic orbits,  and that  its  spectrum
coincides with the average Riemann zeros. This result is generalized  to  Dirichlet $L$-functions using different self-adjoint 
extensions of $H$. We discuss the relation of our work to Polya's fake zeta function and suggest an experimental realization in terms of the Landau model. 
\end{abstract}

%\pacs{02.10.De, 05.45.Mt, 11.10.Hi}  

\maketitle

\vskip 0.2cm

%       DEFINITIONS FOR TEX
%
%%%%%%%%%%%%%%%%%%%%%%%%%%%%%%%%%%%%%%%%%%%%%%%%%%%%%%%%%%%%%%%
%
%
%\def\e{\'e}
%\def\ee{\`e}
%%%%%%%%%%%%%%%%%%%DEFINITIONS%%%%%%%%%%%%%%%%%%%%%%%%%%%%%%%%%
%
\def\oti{{\otimes}}
\def\lb{ \left[ }
\def\rb{ \right]  }
\def\tilde{\widetilde}
\def\bar{\overline}
\def\hat{\widehat}
\def\*{\star}
\def\[{\left[}
\def\]{\right]}
\def\({\left(}      \def\BL{\Bigr(}
\def\){\right)}     \def\BR{\Bigr)}
    \def\BBL{\lb}
    \def\BBR{\rb}
%
%%%%%%%%%%%%%%%%%%%%%%%%%%%%%%%%%%%%%%%%%%%%%%%%%%%%%%%%%%%%%%%
%
\def\zb{{\bar{z} }}
\def\zbar{{\bar{z} }}
\def\frac#1#2{{#1 \over #2}}
\def\inv#1{{1 \over #1}}
\def\half{{1 \over 2}}
\def\d{\partial}
\def\der#1{{\partial \over \partial #1}}
\def\dd#1#2{{\partial #1 \over \partial #2}}
\def\vev#1{\langle #1 \rangle}
\def\ket#1{ | #1 \rangle}
\def\rvac{\hbox{$\vert 0\rangle$}}
\def\lvac{\hbox{$\langle 0 \vert $}}
\def\2pi{\hbox{$2\pi i$}}
\def\e#1{{\rm e}^{^{\textstyle #1}}}
\def\grad#1{\,\nabla\!_{{#1}}\,}
\def\dsl{\raise.15ex\hbox{/}\kern-.57em\partial}
\def\Dsl{\,\raise.15ex\hbox{/}\mkern-.13.5mu D}
%
%%%%%%%%%%%%%%%%%%%%GREEK LETTERS%%%%%%%%%%%%%%%%%%%%%%%%%%%%%%
%
%\def\th{\theta}        \def\Th{\Theta}
\def\ga{\gamma}     \def\Ga{\Gamma}
\def\be{\beta}
\def\al{\alpha}
\def\ep{\epsilon}
\def\vep{\varepsilon}
\def\dep{d}
\def\arc{{\rm Arctan}}
\def\la{\lambda}    \def\La{\Lambda}
\def\de{\delta}     \def\De{\Delta}
\def\om{\omega}     \def\Om{\Omega}
\def\sig{\sigma}    \def\Sig{\Sigma}
\def\vphi{\varphi}
\def\sign{{\rm sign}}
\def\he{\hat{e}}
\def\hf{\hat{f}}
\def\hg{\hat{g}}
\def\ha{\hat{a}}
\def\hb{\hat{b}}
\def\hx{\hat{x}}
\def\hp{\hat{p}}
\def\f{{\bf f}}
\def\g{{\bf g}}
\def\a{{\bf a}}
\def\b{{\bf b}}
\def\fl{{\rm fl}}
\def\sm{{\rm sm}}
\def\QM{{\rm QM}}

%
%%%%%%%%%%%%%%%%%%%CALIGRAPHIC LETTERS%%%%%%%%%%%%%%%%%%%%%%%%%
%
\def\CA{{\cal A}}   \def\CB{{\cal B}}   \def\CC{{\cal C}}
\def\CD{{\cal D}}   \def\CE{{\cal E}}   \def\CF{{\cal F}}
\def\CG{{\cal G}}   \def\CH{{\cal H}}   \def\CI{{\cal J}}
\def\CJ{{\cal J}}   \def\CK{{\cal K}}   \def\CL{{\cal L}}
\def\CM{{\cal M}}   \def\CN{{\cal N}}   \def\CO{{\cal O}}
\def\CP{{\cal P}}   \def\CQ{{\cal Q}}   \def\CR{{\cal R}}
\def\CS{{\cal S}}   \def\CT{{\cal T}}   \def\CU{{\cal U}}
\def\CV{{\cal V}}   \def\CW{{\cal W}}   \def\CX{{\cal X}}
\def\CY{{\cal Y}}   \def\CZ{{\cal Z}}

\def\Hp{{\mathbb{H}^2_+}} 
\def\Hm{{\mathbb{H}^2_-}}

\def\rvac{\hbox{$\vert 0\rangle$}}
\def\lvac{\hbox{$\langle 0 \vert $}}
\def\comm#1#2{ \BBL\ #1\ ,\ #2 \BBR }
\def\2pi{\hbox{$2\pi i$}}
\def\e#1{{\rm e}^{^{\textstyle #1}}}
\def\grad#1{\,\nabla\!_{{#1}}\,}
\def\dsl{\raise.15ex\hbox{/}\kern-.57em\partial}
\def\Dsl{\,\raise.15ex\hbox{/}\mkern-.13.5mu D}
%
%%%%%%%%%%%%%%%%%%%%GREEK LETTERS%%%%%%%%%%%%%%%%%%%%%%%%%%%%%%
%
%%%%%%%%%%%%%%% MATH CHARACTERS %%%%%%%%%%%%%%%%%%%%%%%%%%%%
%
\font\numbers=cmss12
%\font\numbers=cmu10 scaled\magstep1
\font\upright=cmu10 scaled\magstep1
\def\stroke{\vrule height8pt width0.4pt depth-0.1pt}
\def\topfleck{\vrule height8pt width0.5pt depth-5.9pt}
\def\botfleck{\vrule height2pt width0.5pt depth0.1pt}
\def\Zmath{\vcenter{\hbox{\numbers\rlap{\rlap{Z}\kern
0.8pt\topfleck}\kern 2.2pt
                   \rlap Z\kern 6pt\botfleck\kern 1pt}}}
\def\Qmath{\vcenter{\hbox{\upright\rlap{\rlap{Q}\kern
                   3.8pt\stroke}\phantom{Q}}}}
\def\Nmath{\vcenter{\hbox{\upright\rlap{I}\kern 1.7pt N}}}
\def\Cmath{\vcenter{\hbox{\upright\rlap{\rlap{C}\kern
                   3.8pt\stroke}\phantom{C}}}}
\def\Rmath{\vcenter{\hbox{\upright\rlap{I}\kern 1.7pt R}}}
\def\Z{\ifmmode\Zmath\else$\Zmath$\fi}
\def\Q{\ifmmode\Qmath\else$\Qmath$\fi}
\def\N{\ifmmode\Nmath\else$\Nmath$\fi}
\def\C{\ifmmode\Cmath\else$\Cmath$\fi}
\def\R{\ifmmode\Rmath\else$\Rmath$\fi}
%%%%%%%%%%%%%%%%%%%%%%%%%%%%%%%%%%%%%%%%%%%%%%%%%%%%%%%%%%%%%%%%%
 %%%%%%%%%%%%%%%%%% END OF DEFINITIONS %%%%%%%%%%%%%%%%%%%%( c_{\rm sphere} = c_{\rm torus}) %%
 %%%%%%%%%%%%%%%%%%%%%%%%%%%%%%%%%%%%%%%%%%%%%%%%%

\def\barray{\begin{eqnarray}}
\def\earray{\end{eqnarray}}
\def\beq{\begin{equation}}
\def\eeq{\end{equation}}

\def\no{\noindent}

\def\s{\sigma}
\def\Ga{\Gamma}

\def\L{{\cal L}}
\def\g{{\bf g}}
\def\K{{\cal K}}
\def\I{{\cal I}}
\def\M{{\cal M}}
\def\F{{\cal F}}

\def\Im{{\rm Im}}
\def\Re{{\rm Re}}
\def\ti{{\tilde{\phi}}}
\def\tR{{\tilde{R}}}
\def\tS{{\tilde{S}}}
\def\tF{{\tilde{\cal F}}}
\def\om{{\omega_0}} 
\def\oc{{\omega_c}} 
\def\oh{{\omega_h}} 
\def\bm{{\bf m}}
\def\bvS{\vec{{\bf S}}}

%\section{Introduction}

One of the most promising avenues to prove the Riemann hypothesis (RH)  is to find a self-adjoint operator $H$
whose spectrum contains the imaginary part of the non trivial Riemann zeros \cite{E74,T03}. This idea was suggested by  Polya and Hilbert
in the dawn of the XX century and still, one hundred years later,  it  remains unproved, as well as the RH itself 
(see \cite{SH11} for a recent  review on physical approaches to the RH).  There are significant 
hints of the validity of the  Polya-Hilbert conjecture. Two of them are:  
%1) Selberg's trace formula relating the  lengths of  geodesics on a surface with negative curvature and the 
%spectrum of the associated  Laplace-Beltrami operator, 
the Montgomery-Odlyzko law which states,  that 
the local  statistics of the Riemann zeros  is given by the Gaussian Unitary Ensemble (GUE) of Random Matrix Theory, 
and  the formal similarities between counting formulas of  {\em zeros}  in  Number Theory 
and energy levels in Quantum Chaotic systems. In this web  of relationships,  Michael Berry 
suggested   the existence of a classical Hamiltonian whose quantum  version  would realize the Polya-Hilbert  conjecture \cite{B86}. 
This conjectured  Hamiltonian  must satisfy the following   conditions: i) be chaotic, with isolated periodic orbits related to the prime numbers,
ii) break  time reversal symmetry, to agree with the GUE statistics and iii) be quasi-one dimensional. These conditions were derived 
from a formal analogy between the fluctuation part of the Riemann-Mangoldt formula of the zeros of the zeta function
and the Gutzwiller formula for the fluctuation term of the counting of energy levels in a chaotic quantum system. 

In 1999 Berry and Keating showed that the classical Hamiltonian $H_{\rm cl} = xp$ 
fullfills conditions ii)  and iii)   but not condition  i)  \cite{BK99b}.  The  failure of i) is dramatic because this Hamiltonian 
is integrable, and therefore not chaotic,  and moreover the classical trajectories are  not closed, which leads naturally to 
a continuum spectrum. Indeed,  the Hamiltonian
$H_{\cl} = xp$ can be quantized in terms of the self-adjoint operator $\hat{H} = ( x \hat{p} + \hat{p} x)/2$, with $\hat{p} = - i \hbar d/dx$,
and its   spectrum is  given by the  real line \cite{S07a,TM07}. In order to obtain a discrete spectrum,  out of the $xp$ model, 
 Berry and Keating imposed    the conditions $|x| \geq \ell_x$ and
 $|p| \geq \ell_p$, where the minimal length $\ell_x$,  and minimal momentum $\ell_p$ span the Planck area $\ell_x \ell_p = 2 \pi \hbar$
 in phase space.  Subject to these conditions,  a particle with energy $E>0$ 
 describes  a truncated hyperbola in phase space, 
 \beq
 x(t) = \ell_x e^t, \quad p(t) = \frac{E}{ \ell_x} e^{-t}, \quad 0 \leq t \leq T_E =  \log \frac{E}{h}. 
 \label{1}
 \eeq
 The area bounded by this trajectory,  and the $x= \ell_x$ and $p = \ell_p$ axes, measured  in Planck units,  gives
 the semiclassical number of states
 \beq
 N(E) = \frac{E}{2 \pi \hbar} \left( \log \frac{E}{ 2 \pi \hbar} - 1 \right) + \frac{7}{8} + \dots
 \label{2}
 \eeq
 where the constant $7/8$ comes  from a Maslov phase.   
  Rather remarkably, this formula coincides
 with the asymptotic behaviour of the average term  in the Riemann-Mangoldt  formula \cite{E74}, where  $E/\hbar$  is
 interpreted as  the height of a  non trivial zero.   Incidentally, Connes also studied the $xp$ Hamiltonian
 imposing the  constraints  $|x| \leq \Lambda, |p| \leq \Lambda$, where $\Lambda$ is a cutoff \cite{C99}.  In the limit
 $\Lambda \rightarrow \infty$,  one obtains semiclassically a continuum spectrum, where the smooth Riemann zeros appear as 
 missing spectral lines. However,  a more appropiate interpretation of Connes's  result is that Riemann's  formula gives a finite
 size correction to the energy levels. 
  Connes's regularization  were later derived from  the Landau model of a particle moving
 in 2D under the action of external magnetic and electric fields \cite{ST08}. 
 
 A fundamental problem of the Berry-Keating  model is that  the classical trajectories are not closed.  
 The particle starts at the phase space  point $(\ell_x, E/\ell_x)$,  and stops  at the point $(E/\ell_p, \ell_p)$ in a  time $T_E$ (see eq.(\ref{1})). 
The $xp$ hamiltonian breaks time reversal, so the particle cannot return  to its  initial position along the time reversed path. 
Berry and Keating suggested different  ways  to close the trajectories, such as the identification of 
$x$ and $-x$ , and $p$ and $-p$,  or  the use of symmetries, but  no definite conclusion was reached,
and consequently,   the  connection of (\ref{2}) with the Riemann formula could not be put on more solid  grounds. 

\begin{figure}[t!]
\begin{center}
\includegraphics[width=.6 \linewidth]{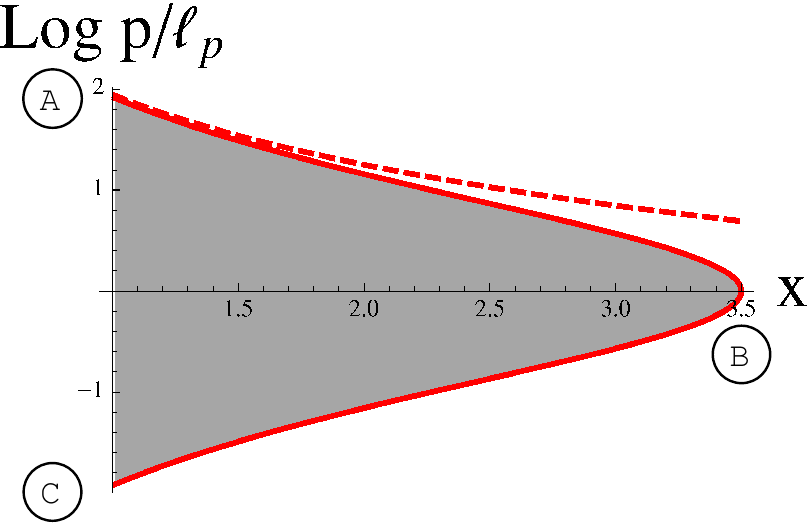}
\end{center}
\caption{
Classical trajectories given in eqs. (\ref{5}) (continuous line) and (\ref{1}) (dotted line)}. 
\label{tray}
\end{figure}

The aim of this letter is to show that the closure problem can be solved by a modification 
of the $xp$ model  that  preserves  several  of its features,  but makes  it  into a consistent quantum model. 
First of all, we shall  constrain  the motion of the particle   to  the half line $ \ell_x \leq x \leq  \infty$ while 
the momentum is allowed to take any real value.  The classical Hamiltonian 
is defined  as 
\beq
H_{\rm cl} = x  \left( p + \frac{ \ell_p^2}{p} \right), \qquad x \geq \ell_x, \quad p \in \Rmath
\label{3}
\eeq
where   $\ell_p$ is a coupling constant with dimensions of  momentum. If $|p| >> \ell_p$, the extra term added to the $xp$ Hamiltonian is negligible, but it becomes
dominant if $|p| <<  \ell_p$,  forbidding  the particle to escape to infinity, since that would cost
an infinite energy. This result is made clear by the solution of  the Hamilton equations 
\beq
\dot{x} = x  \left( 1 - \frac{ \ell_p^2}{ p^2} \right), \quad 
\dot{p} = p + \frac{ \ell_p^2}{p}
\label{4}
\eeq
 given by 
\barray 
x(t) &= & \frac{ \ell_x}{ |p_{0}|}  e^{ 2 t} \sqrt{  ( p_0^2 + \ell_p^2) e^{- 2 t} - \ell_p^2}  \label{5} \\
p(t) & =  & \pm \sqrt{ ( p_0^2 + \ell_p^2)  e^{ - 2 t} - \ell_p^2}.  \nonumber 
\earray 
A complete cycle of a classical trajectory can be described as follows (see fig \ref{tray}). 
The particle starts at the point  $A = (\ell_x, p_0)$ (with $|p_0| \geq  \ell_p$).   Then,  $x$ increases and $p$
decreases monotonically reaching  the turning point $B=(x_{\rm m}(E), \ell_p)$, where  $x_{\rm m}(E) = E/2 \ell_p$ is the maximal
elongation. After that,
the particle moves backwards to  the point $C=(\ell_x, \ell_p^2/p_0)$, which is attained  in a  time
\beq
T_E = \cosh^{-1}   \frac{E}{2 {h}} \rightarrow \log \frac{E}{{h}} \quad  (E >> {h}) 
\label{6}
\eeq
where ${h} \equiv  \ell_x \ell_p$ should not still  be identified with Planck's constant $2 \pi \hbar$. 
At the point $C$,  the particle bounces off, meaning that  its momentum $\ell_p^2/p_0$ becomes  $p_0$, and the
cycle  repeats itself,  with  $T_E$  being  the period. 
The latter process  preserves the total energy, and it is
analogue of the change in the momentum,  $p \rightarrow -p$  of a particle hitting  a wall.  
The classical energies are bounded from below  by the condition $|E|   \geq E_{0}^{\rm cl} = 2 {h}$. The minimum energy correspond
to the static solutions $x= \ell_x$ and $p = \pm \ell_p$. 

An extra condition on the Riemann dynamics is the existence of complex periodic orbits (instantons)
with periods $T_{\rm inst, m} = \pi i  m$ (with $m$ an integer) \cite{BK99b}.  The orbits (\ref{1}) of the $xp$ model
are periodic in imaginary time, but with a wrong  period $2 \pi i$. After a complex  time $ \Delta t = i \pi$, the position and momenta
change sign, which led  Berry and Keating to  suggest  the aforementioned identification between $x$ and $-x$, and $p$ and $-p$, which 
in any case does  not close the orbits.  This problem does not arise  for the Hamiltonian (\ref{3}), which 
 contains complex  periodic  orbits with a period  $\pi i$, as can be seen from  eq.(\ref{5}). 
 
 The semiclassical number of states is given by the phase space area swept  by the particle measured  in units of $2 \pi \hbar$, 
 and it is given by 
\barray 
N(E) & = & 
 \frac{E}{2 \pi \hbar} \left( {\rm cosh}^{-1} \frac{ E}{ 2 {h}} - \sqrt{ 1 - (2 {h}/E )^2 } \right) 
\label{7} \\
& \simeq &  \frac{E}{2 \pi \hbar} \left( \log \frac{E}{ {h}} - 1 \right) + O(E^{-1}),    \quad  \frac{E}{ 2 {h}} >> 1. 
\nonumber 
\earray 
This formula agrees  with eq.(\ref{2}) if $h= 2 \pi \hbar$, up to the constant term, which has not been considered in (\ref{7}). 
Let us now proceed to the quantization of the classical Hamiltonian (\ref{3}). We choose the  normal ordering
prescription, 
\beq
\hat{H} = x^{\frac{1}{2}}    \left( \hat{p} + \frac{ \ell_p^2}{ \hat{p}}  \right) x^{\frac{1}{2}},  
\label{8}
\eeq
where $1/ \hat{p}$ is the 1D Green function satisfying $\hat{p} \;  \hat{p}^{-1}  =\hat{p}^{-1} \hat{p} =  {\bf 1}$, 
and whose matrix elements are 
\beq
\langle x| \,  \frac{1}{\hat{p}} \,  |y \rangle = -  \frac{ i}{ \hbar}    \theta (y-x) 
\label{9}
\eeq
with  $\theta(x)$ the Heaviside step function. $\hat{H}$ acts  on a wave function $\psi$  as
\beq
\hat{H} \psi(x) = - i x^{\frac{1}{2}}  \left[ \hbar   \frac{d}{dx}\left\{ x^{\frac{1}{2}} \psi(x) \right\}  +  \ell_p^2  \int_{\ell_x}^\infty \frac{dy}{\hbar }  \;  
\theta(y-x) y^{ \frac{1}{2}}   \psi(y) \right]. 
\label{10} 
\eeq
This operator is hermitean, i.e. $\langle \psi_1 |  \hat{H}   \psi_2 \rangle =  \langle \hat{H}  \psi_1 | \psi_2 \rangle$, if
both wave functions satisfy the non local boundary condition 
\beq
\hbar  \,  \ell_x^{\frac{1}{2}}  e^{ i \vartheta} \,  \psi(\ell_x) +   \ell_p    \int_{\ell_x}^\infty  dx \, x^{\frac{1}{2}}    \psi(x) = 0.
\label{11}
\eeq
where $\vartheta \in [0, 2 \pi)$. 
To derive (\ref{11}), we have assumed that $\psi(x)$ decays asymptotically   faster that $x^{- 1/2}$. Using eq.(\ref{10}), 
 the Schroedinger equation  $ \hat{H} \psi_E = E \psi_E$
 %
 %\beq
 %\hat{H} \psi_E = E \psi_E
 %\label{12}
 %\eeq
 %
 becomes an  integro-differential equation which can be converted into a second order differential equation
 and a boundary condition. The solution of both  equations yields   a unique square integrable eigenfunction given by  
\beq
\psi_E(x) =  x^{ \frac{i E}{2 \hbar}} \, K_{ \frac{1}{2} - \frac{i E}{2 \hbar}} \left( \frac{ \ell_p x}{\hbar}  \right),  
\label{13}
\eeq
where $K_\nu(x)$ is the  modified K-Bessel function (the normalization factor is not included). 
The asymptotic behaviour of (\ref{13}) is given by 
\beq
\psi_E(x) \sim  \left\{ 
\begin{array}{lc}
x^{- \frac{1}{2} +  \frac{i E}{\hbar}} & x <<  x_{\rm m}(E) \\ 
x^{ - \frac{1}{2}  +  \frac{i E}{ 2 \hbar} } \,  e^{ - \ell_p x/\hbar} & x >> x_{\rm m}(E)  \\
\end{array}
\right. 
\label{14}
\eeq
\begin{figure}[t!]
\begin{center}
\includegraphics[width=.6 \linewidth]{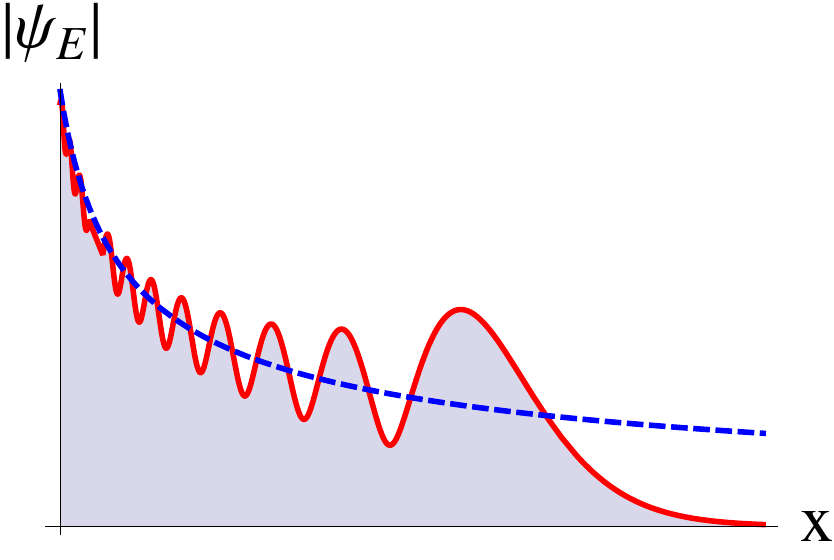}
\end{center}
\caption{
Absolute value wave functions  $\psi_E(x)$, given in eq.(\ref{13}) (continuous line),  and 
$x^{- \frac{1}{2} + \frac{i E}{\hbar}}$ (dotted line). }
\label{psi}
\end{figure}
\noindent 
where $x_{\rm m}(E)$ is the maximal length of the classical trajectory.  
If $x << x_{\rm m}$ the wave function $\psi_E(x)$  behaves, up to  oscilations, as the eigenfunction $x^{- \frac{1}{2} + \frac{i E}{\hbar}}$
of the quantum Hamiltonian $x^{\frac{1}{2}}  \hat{p} x^{\frac{1}{2}}$. 
 However,  $\psi_E(x)$ drops exponentially  in the classical forbidden region (see fig \ref{psi}). 
 The hermiticity of $\hat{H}$, requires the eigenfunctions (\ref{13}) to satisfy the 
  boundary condition (\ref{11}), which in turn provides  the equation for 
  the  eigenenergies, $E_n$, 
\beq
\Xi_{\rm \hat{H}}(E) \equiv  e^{-i  \frac{\vartheta}{2}}
K_{ \frac{1}{2} + \frac{i E}{2 \hbar}} \left( \frac{ h}{\hbar}  \right) + e^{i  \frac{\vartheta}{2}} \,  K_{ \frac{1}{2} - \frac{i E}{2 \hbar}} \left( \frac{ h}{\hbar}  \right)= 0.
\label{15}
\eeq
All the solutions of this equation will  be real,  if the Hamiltonian $\hat{H}$  is,  not only hermitean,  but also self-adjoint.
To verify this property we use the von Neumann theorem:  $\hat{H}$ is a self-adjoint operator 
if the deficiency  indices   $n_+$ and $n_-$  coincide \cite{GP90,AIM05}.  These indices are the number of
linearly independent solutions of the equations $\hat{H}^\dagger  \psi = \pm i \psi$. 
Then  if $n=n_+ = n_-$, the operator   $\hat{H}$ admits infinitely many self-adjoint extensions parameterized by matrices of the unitary group $U(n)$. 
In our case we find that  $n_+= n_-=1$, therefore  the self-adjoint extensions correspond to  a phase,
that  can be identified with the factor   $e^{ i \vartheta}$ appearing   in equations (\ref{11}) and (\ref{15}). 
This ends the proof  of the reality of all the  eigenenergies $E_n$. 

 If $\vartheta \neq \pi$, all the eigenenergies are non vanishing and form time conjugate pairs $\{E_n, - E_{n} \}$ with  their associated eigenfunctions
being related by the time reversal transformation  
 $\psi_{-E_n}(x) = \psi^*_{E_n}(x)$.   If $\vartheta =  \pi$, there is a  unique state of  zero energy $E_0 =0$,  and eigenfunction
 $\psi_{E_0}(x) \propto x^{ - \frac{1}{2}} e^{ - l_p x/\hbar} $, while the non zero energy states form again time conjugate pairs. 
 The ground state energies $\pm E_{0}$ depend strongly on  $\vartheta$ and can be lower or higher  than the classical
 value $E_{0}^{\rm cl}$.

\begin{figure}[t!]
\begin{center}
\includegraphics[width=.49 \linewidth]{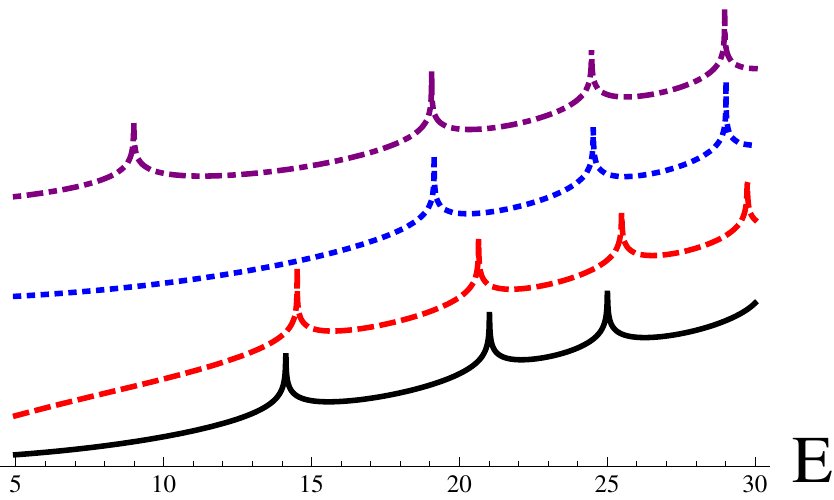}
\includegraphics[width=.49 \linewidth]{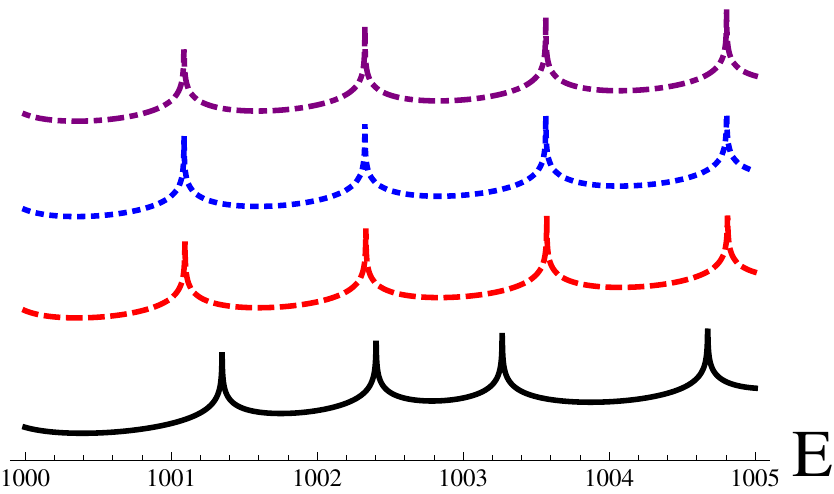}
\end{center}
\caption{
From bottom to top: plot of $- \log|\Xi(E)|$ (Riemann zeros), average Riemann zeros, $- \log|\Xi_{\hat{H}}(E)|$ (eigenenergies of $\hat{H}$
for  $h = 2 \pi \hbar, \vartheta = \pi/4$, 
and $- \log|\Xi^*(E)|$ (Polya zeros). The cusp represents the zeros of the corresponding equations.
 }
\label{spectra}
\end{figure}

To fix  the value of $\vartheta$,  corresponding  to the average Riemann zeros,  we use the asymptotic 
behaviour of eq.(\ref{15}), 
\beq
\Xi_{\hat{H}}(E)  \simeq  \left(\frac{ 4 \pi \hbar}{h} \right)^{\frac{1}{2}} 
 e^{ -  \frac{\pi E}{ 4 \hbar}} \cos \left( \frac{ E}{ 2 \hbar} \log \frac{ E}{ 2 h  e }  - \frac{ \vartheta}{2} \right), 
\label{16}
\eeq 
which vanishes at 
\beq
\frac{ E}{ 2 \pi \hbar} \log \frac{ E}{ 2 h  e }  - \frac{ \vartheta}{2 \pi} = n + \frac{1}{2}, \quad n \in \Zmath. 
\label{17}
\eeq
 If $h= 2 \pi \hbar$
and $\vartheta = 5 \pi/4$, one recovers  the semiclassical estimates for $N(E)$ given in eqs. (\ref{2}) and  (\ref{7}). In references \cite{B86,BKL95},  
it is shown that a better estimate of the average position of the Riemann zeros  is obtained equating   $N(E)$ to a  half integer $n+ \frac{1}{2}$,
rather than  an integer,  which in view of eq.(\ref{17})  yields $\vartheta = \pi/4$ (see fig  \ref{spectra}). 

A confirmation of these results comes from a comparison with Polya's work  on the Riemann $\Xi$-function \cite{P26} (see also \cite{E74,T03}), 
 \barray 
 \Xi(t) & = &  \frac{1}{2} s ( s-1) \pi^{- s/2} \Gamma(s/2) \zeta(s), \; \;  s = \frac{1}{2} + i t , 
 \label{18} 
 \earray
 which is an entire and  even function in $t$, whose zeros coincides with the non trivial zeros of $\zeta(\frac{1}{2} + i t)$. 
 Polya made a Fourier expansion of (\ref{18}) and  truncated it, obtaining 
\barray  
\Xi^*(t) &  =  &  4 \pi^2 ( K_{\frac{9}{4}  +  \frac{i t}{2}}( 2 \pi) + K_{\frac{9}{4}  -  \frac{i  t}{2}}( 2 \pi)), 
\label{19}
 \earray
which  is called Polya's  {\em fake} zeta function.  since it shares several  properties with $\Xi(t)$. 
First of all, the zeros of $\Xi^*(t)$ and  $\Xi(t)$, agree in average,  as can be seen using  the asymptotic expansion \cite{T03}. 
 \barray  
\Xi^*(t) &  \sim   &  \pi^{\frac{1}{4}}   2^{- \frac{5}{4}} t^{ \frac{7}{4}} e^{ - \frac{ \pi t}{4}} 
 \cos \left( \frac{ t}{ 2} \log \frac{ t}{ 2 \pi   e }  + \frac{ 7 \pi}{8} \right). 
\label{20}
 \earray
This expression vanishes when the argument of the cosine is $n + \frac{1}{2}$, which confirms  the aforementioned
rule for the average location of the  Riemann zeros, and in turn the choice  $\vartheta = \pi/4$. 
A more  remarkable fact is that {\em all}  the
zeros   of  $\Xi^*(t)$  are real, as was proved  by Polya using a general  theorem on entire functions
 \cite{P26}. This  theorem can also be applied to prove the reality of all the zeros of 
 $\Xi_{\hat{H}}(E)$, a result  that we obtained using the 
  self-adjointness of the operator $\hat{H}$. 
 
The RH is a particular case of the 
 generalized Riemann hypothesis (GRH), which asserts that all  the non trivial zeros of the 
 Dirichlet $L(\chi,s)$-functions, associated  to the Dirichlet character $\chi$,   lie on the critical line $\Re \, s= \frac{1}{2}$. These functions 
 are defined by a series and associated Euler product   ($\Re \, s > 1$)  
 \beq
L(s, \chi) = \sum_{n=1}^\infty \frac{ \chi(n)}{ n^s} = \prod_{p: {\rm prime}} \frac{1}{ 1 - \chi(p) p^{-s}}, 
\label{21} 
\eeq
and their  analytic extension to the  complex plane.  $\chi(n)$ are  multiplicative arithmetic  functions, i.e. 
 $\chi(nm) = \chi(n) \chi(m), \chi(n + q m) = \chi(n), \chi(1) = 1$, where $q$ is the modulus of $\chi$. 
%The Riemann zeta function corresponds to the trivial character $\chi(n) = 1, \forall n$. 
$L$-functions associated to primitive characters satisfy 
the functional   relation   \cite{D80}, 
\beq
\xi(s, \chi) = \left( \frac{ \pi}{ q} \right)^{ - \frac{ s+a_\chi}{2} } \Gamma \left( \frac{ s + a_\chi}{2} \right)   L(s, \chi) = \epsilon_\chi \, 
\xi(1-s, \bar{\chi})
\label{22}
\eeq
where $a_\chi$ is the {\em parity} and $\epsilon_\chi$ is the sign of a Gaussian sum,
\beq
a_\chi = \frac{1-  \chi(-1)}{2}, \; 
\epsilon_\chi= \frac{ \tau_\chi}{ i^{a_\chi}  \,  q^{1/2}}, \,  \tau_\chi= \sum_{n=1}^q \chi(n) \, e^{ \frac{ 2 \pi i n}{q}} 
\label{23}
\eeq
A $L$-function is  even (odd) if $a_\chi=0 \,  (1)$. The Riemann zeta function corresponds to the trivial character $\chi(n) = 1, \forall n$, 
with $a_\chi=0, \epsilon_\chi=1$.  Equation (\ref{22}) yields the average location of the zeros of $L(\chi, s)$ 

\beq
 \frac{t}{2  \pi } \log \frac{ q \,  t}{2 \pi e } -  \frac{1}{8}  + \frac{a_\chi +\epsilon_\chi -1 }{4} = n + \frac{1}{2} 
\label{24}
\eeq
which leads us to the following identification of parameters in the $\hat{H}$ model (see eq.(\ref{17})), 
\beq
 \frac{E}{\hbar} = t,  \quad 
h  = \frac{ 2 \pi \hbar}{ q}, \qquad \vartheta =  \frac{\pi}{4} ( 3  - 2 a_\chi- 2 \epsilon_\chi). 
\label{25}
\eeq
The Riemann zeta function corresponds to the case $q=1$,  for which  $h= 2 \pi \hbar, \vartheta = \pi/4$. The correspondence
(\ref{25}) implies that the constant $h$ is quantized as a function of the modulus of the $L$-functions, 
attaining  the classical limit,  $h \rightarrow 0$, when $q \rightarrow \infty$. 

A physical realization of the Hamiltonian (\ref{3}) is suggested by the work of reference \cite{ST08}, 
which showed  that $H_{\rm cl}= xp$  emerges as the effective Hamiltonian of an electron
moving in the $x-y$ plane,  subject to the action of a uniform magnetic field $B$, perpendicular to the plane,  and an electrostatic potential 
$V(x,y) = V_0 \,  x y$.  If $V=0$, the electron occupies  the lowest Landau level which is completely
degenerate.  This degeneracy  is broken by the potential $V(x,y)$,  which in  perturbation theory  becomes 
the   1D Hamiltonian 
$H_{\rm eff} = \omega_0 \,  x p$, where $\omega_0 = V_0 \ell^2/\hbar$  ( $\ell= \sqrt{ \hbar c/eB}$ is the magnetic length). 
The latter Hamiltonian is  obtained  replacing $y \rightarrow \ell^2 p/\hbar$ in $V(x,y)$.  
Consider now that the particle moves in the half-plane $x \geq \ell$ and that the electrostatic  potential is 
\beq
V(x,y) = V_0 x \left(  y + \frac{ ( 2 \pi \ell /q)^2}{ y} \right).   
\eeq
Then, the effective Hamiltonian,  in the lowest Landau level, in units of  $\omega_0$,  becomes  (\ref{3}), with the identifications
$\ell_x = \ell, \ell_p = 2 \pi \hbar/q \ell$ and $h = 2 \pi \hbar/q$. We expect the parameter $\vartheta$ to arise from 
an electric field applied at the boundary $x= \ell$ of the system.

In summary, we have reformulated the Berry-Keating $xp$ model in terms  of  the classical Hamiltonian
$H_{\rm cl} = x( p + \ell_p^2/p)$ defined on the half-line $x \geq \ell_x$, which posseses    closed orbits  
 and whose  semiclassical spectrum agrees
with the average Riemann zeros. The quantization of  this Hamiltonian, yields a
self-adjoint operator $\hat{H}$, and a non local boundary condition parameterized by an angle $\vartheta$. 
The  spectrum of $\hat{H}$ agrees asymptotically with the semiclassical result and the eigenenergy equation 
 is similar  to  Polya's fake zeta function that  approximates the Riemann's $\Xi$ function. 
The construction is generalized to the  Dirichlet $L-$ functions,   supporting  the idea that the GRH could have a proof based on 
a common quantum mechanical  model.  To achieve this goal one has of course to find the quantum origin of the fluctuations
of the Riemann zeros. This work suggest two possible scenarios. One is to discretize the dynamics as in the Arnold's cat map.
The other is to modify the Hamiltonian in a non trivial way. Further research is required to clarify which path is the best.

{\bf Acknowledgements.- }  We are   grateful to Paul  Townsend, Michael  Berry and Jon  Keating for conversations.  
This work has  been financed  by Ministerio de Educaci\'on y Ciencia, Spain (grant FIS2009-11654) and  Comunidad de Madrid (grant QUITEMAD).

\end{document}